\documentclass[runningheads]{llncs}
\usepackage[T1]{fontenc}
\usepackage{booktabs}
\usepackage{graphicx}
\usepackage{braket}
\usepackage{qcircuit}

\usepackage[bookmarks,bookmarksopen,bookmarksdepth=3, hidelinks, colorlinks=false]{hyperref}

\usepackage{soul}
\usepackage{braket}
\usepackage{amsmath}
\usepackage{amsfonts}

\usepackage{color, colortbl}
\definecolor{Gray}{gray}{0.9}

\definecolor{LightCyan}{rgb}{0.88,1,1}
\usepackage{subcaption}
\begin{document}
% \raggedbottom
% \setlength{\parskip}{0pt}
%
% \title{Quantum Cryptanalysis Landscape \\ of Shor’s Algorithm \\ for Elliptic Curve Discrete Logarithm Problem\thanks{This work was supported by Institute for Information \& Communications Technology Planning \& Evaluation (IITP) grant funded by the Korea government (MSIT) (No.2019-0-00033, Study on Quantum Security Evaluation of Cryptography based on Computational Quantum Complexity, 50\%) and this work was also supported by Institute of Information \& Communications Technology Planning \& Evaluation (IITP) grant funded by the Korea government (MSIT) (2019-0-01343, Regional strategic industry convergence security core talent training business).}}
% %
% \title{On Evaluating Quantum Point Doubling Circuit in Shor’s Algorithm for Binary Elliptic Curves} %for paper aja
\title{Quantum Circuit Designs of Point Doubling Operation for Binary Elliptic Curves}
%

% \titlerunning{On Evaluating Quantum Point Doubling Circuit in Shor’s Algorithm...}
\titlerunning{Quantum Circuit Designs of Point Doubling for Binary Elliptic Curves}

% % an abbreviated paper title here
% %
% \author{Harashta Tatimma Larasati\inst{1,2}\orcidID{0000-0001-6143-4134} \and \\
% Howon Kim\inst{1}\orcidID{0000-0001-8475-7294}}
\author{Harashta Tatimma Larasati \and 
Howon Kim}
% %
\authorrunning{H.T. Larasati and H. Kim}
% \institute{School of Computer Science and Engineering, Pusan National University, Busan 609735, Republic of Korea \\
% \email{\{harashta, howonkim\}@pusan.ac.kr}%\\ \and School of Electrical Engineering and Informatics, Institut Teknologi Bandung, Bandung 40116, Indonesia}
% % \email{harashta@stei.itb.ac.id}

%
\maketitle              % typeset the header of the contribution
%
% \input{0.abstract}
% \input{1.intro}
% \input{2.prelim}
% \input{3.landscape}
% % \includepdf[pages=-]{mappingtabelexcel.pdf}
% % \includepdf[pages=-]{0624_fix_fromlate_abbrev_resizedA4.pdf} %nanti dijadiin 2 tabel aja? 
% \input{4.review}
% \input{5.conclusion}

\begin{abstract}%percantik later
In the past years, research on Shor’s algorithm for solving elliptic curves for discrete logarithm problems (Shor’s ECDLP), the basis for cracking elliptic curve-based cryptosystems (ECC), has started to garner more significant interest. To achieve this, most works focus on quantum point addition subroutines to realize the double scalar multiplication circuit, an essential part of Shor’s ECDLP, whereas the point doubling subroutines are often overlooked. In this paper, we investigate the quantum point doubling circuit for the stricter assumption of Shor’s algorithm when doubling a point should also be taken into consideration. In particular, we analyze the challenges on implementing the circuit and provide the solution. Subsequently, we design and optimize the corresponding quantum circuit, and analyze the high-level quantum resource cost of the circuit. Additionally, we discuss the implications of our findings, including the concerns for its integration with point addition for a complete double scalar multiplication circuit and the potential opportunities resulting from its implementation. Our work lays the foundation for further evaluation of Shor’s ECDLP.
\end{abstract}

%%%%%%%%%%%%%%%%%%%%%%%%%%%%%%%%%%%%%%%%%%%%%%%%%%
\section{Introduction}
% In the past years, research on Shor’s algorithm for solving elliptic curves for discrete logarithm problems (Shor’s ECDLP) \cite{shor1994algorithms,shor1999polynomial} has started to garner more significant attention. 
Over the decade, there has been a growing interest in Shor’s algorithm for solving the elliptic curve discrete logarithm problems (i.e., Shor's ECDLP) \cite{shor1994algorithms,shor1999polynomial}.
Acknowledged to render existing elliptic curve-based cryptosystems (ECC) breakable in polynomial time \cite{roetteler2017quantum}, this algorithm has the potential to accomplish its objective of cracking existing public-key cryptography (PKC) sooner than its more popular counterpart, i.e., Shor’s factoring algorithm for cracking RSA, due to its lower quantum resource requirement for the same security level \cite{kirsch2015quantum,proos2003shor}. In particular, the advantage of lower key size in ECC is \textemdash ironically \textemdash  the reason why it is in graver danger in the presence of a quantum computer, considering the current development of quantum computing that is still in the early stage, which often favors the number of qubits as the most essential metric.

To date, several works have discussed how to concretely realize Shor’s ECDLP for quantum cryptanalysis purposes \cite{roetteler2017quantum,haner2020improved,banegas2021concrete,gouzien2023computing,liu2022quantum}, with heavily referenced state-of-the-art works \cite{roetteler2017quantum,haner2020improved,banegas2021concrete} primarily assessing the implementation for the superconducting qubits architecture as arguably the most prominent quantum hardware platform. Starting from the works by Roetteler et al. \cite{roetteler2017quantum} and perfected by Haner et al. \cite{haner2020improved}, which both consider prime curves implementation, the landscape then extends to binary elliptic curves by Banegas et al. \cite{banegas2021concrete},
% with the metrics 
%% WTD: as \hl{presented in Table} [1]. %kl ga keburu, close aja di sini.

All those advancements are based on the pioneering efforts of Proos and Zalka \cite{proos2003shor}, one of the earliest works to translate the high-level Shor’s ECDLP algorithm into the description of their possible quantum circuit derivation. Over time, their paper has established itself as the standard reference for subsequent papers in the literature that aims to optimize the quantum circuit implementation of Shor’s ECDLP, which has been made easier for testing, verification, and concretely estimating the quantum resource requirement by leveraging reversible circuit and quantum computing simulators that have emerged in the past decade (e.g., RevKit, LIQ$Ui\ket{}$, and the more recent ProjectQ, Qiskit, Microsoft QDK/Azure Quantum, and Q-Crypton). 
% %%WTD: reference (e.g., RevKit \cite{}, LIQ$Ui\ket{}$ \cite{}, and the more recent platforms such as ProjectQ \cite{}, Qiskit \cite{}, Microsoft QDK or Microsoft Azure Quantum \cite{}, and Q-Crypton \cite{}). 

\begin{figure}[tb!]
% {.5\textwidth}
\centering
\includegraphics[width=.8\textwidth]{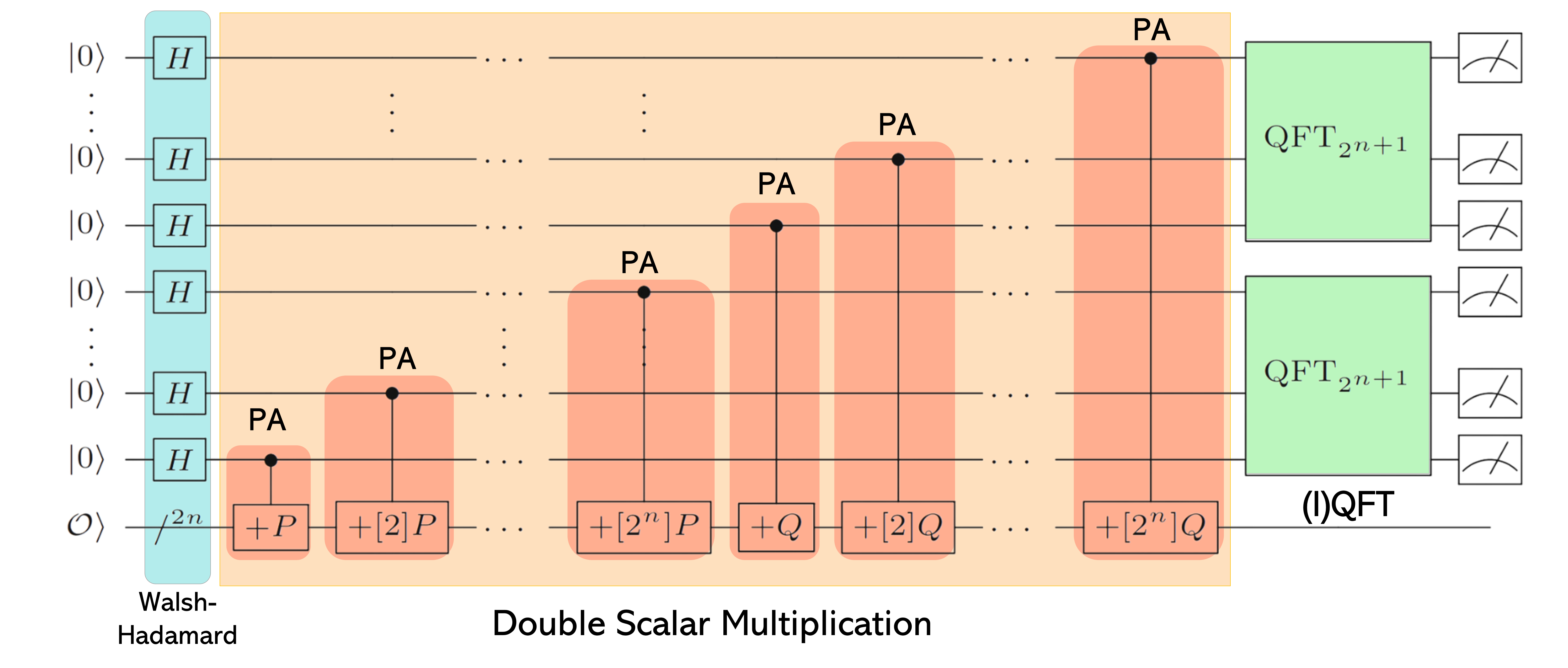} %nanti ganti warna aja, skrg persis aja dulu gpp
\caption{Quantum circuit of Shor's algorithm for solving the elliptic curve discrete logarithm problem (ECDLP). Figures adapted from \cite{roetteler2017quantum,larasati2023depth}.\label{fig:shor_ecdlp}}
\end{figure}

From our observation, these papers preserve the scope provided by Proos and Zalka \cite{proos2003shor}. That is, for cracking ECC via Shor’s ECDLP, the rule can be simplified by considering only the generic case (i.e., for points $P + R$ where $P, R \neq O$, and $P \neq \pm R$) for the elliptic curve group operation \cite{proos2003shor}. In other words, to achieve the \textit{double scalar multiplication}, the essential components in Shor’s ECDLP circuit (see Fig. \ref{fig:shor_ecdlp}), computation will be done solely by a series of \textit{point addition} operations. Meanwhile, the other operation to perform a more special case where $P = R$, namely the \textit{point doubling} operation, is set aside. The authors of \cite{proos2003shor} argued that the expected loss of fidelity from the absence of this operation would still be negligible, which was also agreed upon by succeeding papers, e.g., \cite{kaye2004optimized}.

Nevertheless, when considering the stricter assumption where the occurrence of $P = R$ is more probable during computation and minimum fidelity loss is expected from the construction, point doubling operation will also hold considerable significance. In this case, exploring the point doubling operation, including its quantum circuit construction and the analysis of its quantum resource, will be very beneficial and insightful for more precise resource estimation of Shor's ECDLP.

In this study, we examine the point doubling operation as required for the less relaxed case of Shor’s ECDLP, i.e., when the elliptic curve points happen to be the same two points. To the best of our knowledge, this subject, including the possible quantum circuit implementation, has so far been absent in state-of-the-art works in quantum cryptanalysis. For this initial work, we focus on point doubling circuit for binary elliptic curves, whose inherent characteristics make it simpler for tinkering and constructing the operation compared to the prime curves counterpart. To highlight our contributions, we start by analyzing the point-doubling formula and identifying the challenges in its construction with their possible solution. Subsequently, we design the quantum circuits for elliptic curve point doubling to suit several scenarios and analyze its quantum resource cost in a high-level view. Furthermore, we also provide a more detailed discussion of the aspects related to prime curves and the concerns when incorporating the circuit for use in Shor’s algorithm. %aftermath and elaborate %describe the next steps to incorporate the circuit for Shor’s algorithm. 

The contribution of this paper can be summarized as follows:
\begin{itemize}
    \item We examine the elliptic curve point doubling operation, which is rarely explored in literature. In particular, we discuss the challenges, analyze the formula and the implementation possibility of point doubling circuits for binary elliptic curves.
    \item We design the corresponding quantum circuit, incorporate several optimization and address the uncomputation, then analyze the high-level quantum resource cost of the circuit.
    \item We provide an in-depth discussion of our findings and other aspects relevant to point doubling, the concerns when incorporating the circuit with point addition for a complete double scalar multiplication circuit, as well as the open possibilities arising from point doubling implementation.
\end{itemize}

%%WTD:% \hl{STEPS to describe blm ada lho..hayo sesuaikan si Intro, contrib \& Conclusion kl emang gaada}

% The rest of this paper \hl{is structured as follows.}

%%%%%%%%%%%%%%%%%%%%%%%%%%%%%%%%%%%%%%%%%%%%%%%%%%
\section{Preliminaries}

\subsection{Shor’s ECDLP}  %baru parafrase gpt dr flt lalu kuedit
The security of elliptic curve cryptography (ECC) is based on the hardness of the elliptic curve discrete logarithm problem (ECDLP). In this problem, given two points $P$ and $Q$ on an elliptic curve of order $r$, it is easy to compute the point multiplication $Q = kP$ when the scalar $k$ and the base point $P$ are known. In contrast, the reverse problem of finding the scalar $k$ given both points $P$ and $Q$ is computationally intensive \cite{roetteler2017quantum} and considered classically intractable. 

\textbf{How it works.} Shor's algorithm for solving elliptic curve discrete logarithm problems (Shor's ECDLP) works by essentially running a brute-force attack of computing the scalar multiplication of all states, but intelligently utilizing quantum interference to boost the likelihood of obtaining the desired result while suppressing the undesired value via quantum Fourier transform (QFT). As illustrated in Fig. \ref{fig:shor_ecdlp}, the algorithm consists of three registers with two $n+1$-sized quantum registers initialized in the state $\ket{0}$ appended with the Walsh-Hadamard (i.e., Hadamard gate on each qubit), which yields the state $\frac{1}{2^{n+1}} \sum_{k,l=0}^{2^{n+1}-1}{{k,l}}$. Subsequently, conditional to the state of the register containing $k$ or $l$, the corresponding multiple of points $P$ and $Q$ are added via the double-scalar multiplication circuit, performing the mapping as in Eq. \ref{eq:shor_ecdlp_formula} \cite{roetteler2017quantum}, 

\begin{equation}
\frac{1}{2^{n+1}} \sum_{k,l=0}^{2^{n+1}-1}{{k,l}}\mapsto \sum_{k,l=0}^{2^{n+1}-1}{{k,l}}\ket{\left[k\right]P + \left[l\right]Q}
\label{eq:shor_ecdlp_formula}
\end{equation}

\noindent
before appending QFT and measuring the result. Finally, classical post-processing is performed, which theoretically can yield the sought value with high probability. Consequently, this algorithm enables an adversary with a large-scale, full-fledged quantum computer to obtain $k$ by running the algorithm a few times. 

%%for journal: elaboration on HOW TO MAKE ECPM, mainly pk PA, kenapa ga bnanyak PD/argumen2 terkait (cek goodocs outline)

\textbf{Quantum scalar multiplication circuit.} In existing works, as previously shown in Fig. \ref{fig:shor_ecdlp}, the quantum (double) scalar multiplication circuit comprises solely of (controlled) point addition operation, simplifying the operation by making the added point fixed. However, this does not cover the case where both points are the same, which necessitate the use of point doubling operation, therefore may yield incorrect result when doubling the points \cite{proos2003shor}. Even though it is argued that the fidelity loss from this is small, the stricter case will require the analysis of the point-doubling circuit as well. Therefore, it is beneficial to analyze the point doubling circuit, which we start with this paper.

% . , as shown  has been proposed For achieving scalar multiplication circuit, existing works employ point addition circuit 

% \subsection{Binary Elliptic Curves} %masih dari FLT no paraphrase. GANTI PAS LNCS PROCEEDING AJA!

% For the binary elliptic curves case, the point addition circuit is as shown in Fig. \ref{fig:ecc_pa}. 

% Binary elliptic curves PA and PD 
% Quantum PA (M denotes multiplication, etc
% Division via inversion plus multiplication
% Addition by a constant
% S squarings)
% Related work mentioning PD
% MASUKIN BAHASAN COST DIKIT, ky conclusions/no of operation di result! misal, sekian no of operation...which necessitate multi control (MCT)...jd gabisa ngitung langsung, leave for future work.

\subsection{Binary Elliptic Curves in the Quantum Realm} %PERSIS AMBIL DARI FLT
From a quantum cryptanalysis perspective, an ordinary binary elliptic curve is often considered instead of other stronger variants such as supersingular \cite{banegas2021concrete}. Here, we first describe the theoretical concept of binary elliptic curves. The Weierstrass equation for an ordinary binary elliptic curve is described in Eq. \ref{eq:binaryecc_eq_weierstrass}, 
\begin{equation}\label{eq:binaryecc_eq_weierstrass}
    y^2 + xy = x^3 + ax^2 +b
\end{equation}
where $a \in \mathbb{F}_2$ and $b \in \mathbb{F^*}_{2^m}$ (i.e., the extension field). Then, the points on this elliptic curve, $P = (x,y) \in \mathbb{F^2}_{2^m}$, form a set of points that can be computed under the elliptic curve group law comprising \textit{point addition} and \textit{point doubling} operations. In particular, point addition, e.g., $P_1 + P_2 = P_3$, with $P_1 = (x_1, y_1)$, $P_2 = (x_2, y_2) \neq \pm P_1$, and $P_3 = (x_3, y_3)$, can be computed by following Eqs. \ref{eq:pa_x3} to \ref{eq:pa_lambda}. 
\begin{align}
x_3 & = \lambda^2 + \lambda + x_1 + x_2 + a \label{eq:pa_x3} \\
y_3 & = \lambda (x_1 + x_3) + x_3 + y_1 \label{eq:pa_y3} \\
\lambda & =  \frac{y_1 + y_2}{x_1 + x_2} \label{eq:pa_lambda}
\end{align}
Meanwhile, the point doubling calculation is as shown in Eqs. \ref{eq:pd_x3} to \ref{eq:pd_lambda}.  \cite{hankerson2006guide,pornin2022efficient}.

\begin{align}
x_3 & = \lambda^2 + \lambda + a = {x_1}^2 + \frac{b}{{x_1}^2} \label{eq:pd_x3} \\
y_3 & = {x_1}^2 + (\lambda + 1) x_3 \label{eq:pd_y3} \\
\lambda & = x_1 + y_1/x_1 \label{eq:pd_lambda}
\end{align}

\begin{figure}[!tb] 
\centering
\resizebox{\linewidth}{!}{
\scalebox{1.5}{
\Qcircuit @C=1em @R=.7em {
&&\lstick{\ket{x_1}} & {{/}^{n}}\qw & \gate{+x_2} & \ctrl{2} & \ctrl{2} & \gate{+a +x_2} & \targ & \targ & \push{\rule{0em}{1.2em}}\qw & \ctrl{2} & \ctrl{2} & \gate{+x_2} & \ctrl{1}  &\rstick{\ket{x_3} or \ket{x_1}}\qw &&&&& \\
&& \lstick{\ket{q}} & \qw & \ctrl{1} & \qw & \qw & \ctrl{-1} & \ctrl{-1} & \ctrl{-1} & \qw & \qw & \push{\rule{0em}{1.2em}}\qw & \ctrl{1} & \ctrl{1}  &\rstick{\ket{q}}\qw \\
&& \lstick{\ket{y_1}} & {{/}^{n}}\qw & \gate{+y_2} & \ctrl{1} & \gate{M} & \gate{S} & \ctrl{-1}  & \push{\rule{0em}{1.2em}}\qw & \gate{S} & \gate{M} & \ctrl{1} & \gate{+y_2} & \targ  &\rstick{\ket{y_3} or \ket{y_1}}\qw \\
&& \lstick{\ket{0}} & {{/}^{n}}\qw & \qw & \gate{D} & \ctrl{-1} & \ctrl{-1} & \qw & \ctrl{-2} & \ctrl{-1} & \ctrl{-1} & \gate{D}  & \qw & \push{\rule{0em}{1.2em}}\qw &\rstick{\ket{0}}\qw}
}
}
\caption{Point addition circuit for binary elliptic curves proposed by Banegas et al \cite{banegas2021concrete}.} \label{fig:ecc_pa}
\end{figure}
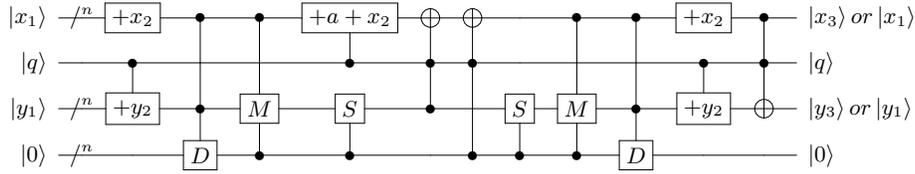

\textbf{Constructing the quantum circuit.} From the group law formula above, the corresponding quantum circuit can be constructed.
\footnote{All classical computation can be simulated on a quantum computer by reversible gates, e.g., Toffoli (the most common), Fredkin, or Barenco gates \cite{williams2010explorations}. %, see Fig. \ref{fig:universal_gates}. 
However, how to efficiently perform the operation is a whole different topic pursued by researchers.}
Regarding the quantum point addition circuit, the recent concrete construction is by Banegas et al. \cite{banegas2021concrete}, which is presented in Fig. \ref{fig:ecc_pa}. As inferred from the figure, the circuit requires three registers of size $n$ in which two serve as input/output registers and one as a clean ancilla register, plus one qubit serving as the control \textemdash which in the full scheme of Shor's ECDLP circuit will be associated with the qubit in the upper registers (the ones appended by Walsh-Hadamard). Additionally, the circuit utilizes two multiplications, two divisions, and two squarings \textemdash all of which are conditionally controlled, linked to the control qubit and other associated register \textemdash and several (controlled) additions and addition by a constant. 

\textbf{Quantum resource cost.} In terms of the exact resource count, however, it will greatly depend on the underlying subroutines employed since the aforementioned circuit is still a high-level architecture that will be broken down into its finer-grained components. For instance, choosing to use between two different inversion techniques: greatest common divisor (GCD) \cite{banegas2021concrete} or Fermat's Little Theorem (FLT) \cite{banegas2021concrete,larasati2023depth,taguchi2023concrete} for the division subroutines, or between Schoolbook \cite{vedral1996quantum} and Karatsuba multiplication \cite{vanhoof2019space,putranto2023depth,gidney2019asymptotically,jang2022optimized}  will yield quite different performance metrics, including in terms of the total number of qubits (i.e., qubit count or circuit width), circuit depth (i.e., the longest path for the quantum operations to run on the quantum hardware, gate count (i.e., the total number of quantum gates), as well as the more specific terms like Toffoli depth and Toffoli count \cite{gyongyosi2020circuit,ucberkeley_2007}.

\vspace{-1.5em}

%%%%%%%%%%%%%%%%%%%%%%%%%%%%%%%%%%%%%%%%%%%%%%%%%%
% \section{Challenges for Implementing Point Doubling}
% \section{Evaluating Point Doubling Circuit for ...}
\section{Quantum Circuit Designs of Point Doubling Operation for Binary Elliptic Curves}
In this section, we start by elaborating on the challenges in constructing point-doubling operations. Furthermore, we provide three circuits for point doubling to suit different design considerations. In this work, we aim to be clear also for non-expert audiences; therefore, we describe our thought process to develop the resulting circuit.

\subsection{Challenges on Quantum Point Doubling Construction}
Before going into detail about the point-doubling circuit itself, it would be better to start with the differences between point addition and point doubling from the quantum perspective that we are able to identify. Constructing a point-doubling circuit poses relatively more difficulties than a point-addition circuit. Firstly, to implement point addition, previous works \cite{proos2003shor} proposed simplification by making one of the two points constant, which is added conditionally depending on the state of the control qubit (which represents each qubit that is appended by Hadamard gates in the upper registers of Shor’s ECDLP, see Fig. \ref{fig:shor_ecdlp}). %that connects to the respective Hadamard gates in the input register (see Fig 1)

With this, the point to be added (i.e., $P_2(x_2, y_2)$) is appended conditionally as a constant; hence can be pre-set and precomputed classically. Furthermore, by making the point a constant, the uncomputation process can be performed with ease since the added point can be immediately subtracted or uncomputed as soon as they are no longer needed in the calculation, making it practical and more efficient. Secondly, as mentioned in \cite{proos2003shor} and further elaborated in \cite{roetteler2017quantum}, by looking further at the point addition formulas (Eqs. \ref{eq:pa_x3} to \ref{eq:pa_lambda}), the value of $\lambda$ in point addition has a direct, clear relation with both $x_1$ and $x_3$, as well as $y_1$ and $y_3$ (i.e., $x_3$ can be obtained from appending $x_2$ and $\lambda$ to $x_1$ with other relevant operations (Eq. \ref{eq:pa_x3}), and similarly, $y_3$ can be obtained from appending $y_1$ and $\lambda$ to $y_1$ with other relevant operations (Eq. \ref{eq:pa_y3})). Here, we say that the initial state of $x_1$ and $y_1$ can be "consumed" to obtain the final desired computation. Then, by intelligently arranging the circuit, we can straightforwardly transform the initial state ($x_1, y_1$) to the subsequent state ($x_3, y_3$). As a result, an efficient computation (and uncomputation) can be achieved, and a clear reversibility relationship can be maintained.

%%made clearer by GPT from my own writing
On the other hand, the construction of point doubling is relatively tricky. First, we will discuss point doubling in a broader view without restricting ourselves to the case of Shor’s algorithm requirement. 
In the case of point doubling, both points involved in the computation share identical values ($P = R$). Unlike point addition, where it is reasonable to assume that the second point is constant and its value is known in advance, the same assumption does not hold for point doubling. Intuitively, if the point to be doubled were known beforehand, then the whole point doubling operation would serve no purpose. % since   because if it is, then there is no use of that.
%, rendering the assumption ineffective.

Hence, the practice of appending the value of the second point, as seen in point addition, is not applicable in this scenario. Consequently, an extra placeholder (register) will be required to store or append the same point, which can be achieved through a "fan-out" or "copy" operation using CNOT gates. Moreover, due to both input points being quantum and the operations being conditionally dependent on the state of a controlled qubit $q$, many of the operations will ultimately require "elevation,": CNOT becomes CCNOT (controlled-controlled NOT gate a.k.a. Toffoli gate), CCNOT becomes CCCNOT (multi-controlled Toffoli gate), and so on, leading to more complex operations.

% % For point doubling, both points to be computed share identical values ($P = R$). Unlike point addition that would still be sensible to assume the second point is constant and the value is known, intuitively, the assumption will not work in the point doubling case (meaning that the point would not be known beforehand). 
% Thus, we cannot append the value of the second point as done on point addition. Consequently, this case will require an additional placeholder (register) for containing or appending the same point (which can be performed by a “fan-out” or “copy” operation via CNOT gates). Additionally, since both input points are quantum and the operations are still conditional to the state of controlled qubit q, many of the operations will ultimately need to be “elevated”: CNOT becomes CCNOT (Toffoli), CCNOT becomes CCCNOT (multi-control Toffoli gates), and so on, which results in more complex operations.

Furthermore, examining the point doubling formula in Eqs. \ref{eq:pd_x3} to \ref{eq:pd_lambda}, obtaining $x_3$ from $x_1$ and $y_3$ from $y_1$, is not as straightforward. The term $x_1$ does not directly evolve into $x_3$, and similarly for $y_1$ and $y_3$. In detail, as inferred from Eq. \ref{eq:pd_x3}, obtaining $x_3$ from $x_1$ requires "copying" $x_1$ to be squared and then appended (i.e., $x_1^2 + \frac{b}{x_1^2})$, while obtaining it from $\lambda$ does not require any $x_1$. Hence, we say that it does not "consume" the initial state. Similarly for $y_3$ as obtaining it does not make use of $y_1$ at all. As a consequence, the initial value of $y_1$ may need to be preserved in the circuit as it can not be erased, hence requiring a placeholder (such as an ancilla register) to hold its value. This makes it challenging to devise an efficient design for its quantum circuit implementation. % \hl{LANJUTKAN/ganti}.

\subsection{Proposed Quantum Circuits}
%%WTD: \hl{bikin table komponen2 yg bisa dipake? (inversion, mult, squaring, multicontrol) bikin tabelnya aja? mm}
Despite the challenges, there are still opportunities from the point-doubling formula that we can leverage to implement the circuit rather efficiently. We observe that there exists an indirect relation that can be taken advantage of. In particular, notice that $x_1$ has a direct relation to $y_3$, while $y_1$ has a direct relation to $x_3$. By utilizing this correlation, it is possible to transform $x_1$ and $y_1$ into $y_3$ and $x_3$, respectively. Thereby, a relatively efficient circuit can still be obtained, albeit with a "twisted" input-output relation (i.e., where $x_1$ maps to $y_3$ and $y_1$ maps to $x_3$ instead of the aligned mapping of $x_1$ to $x_3$ and $y_1$ to $y_3$).

% notice that x1 has a direct relation to y3, whereas y1 has a direct relation to x3. This correlation can be used to transform x1 and y1 to y3 and x3, respectively. Thereby, a rather efficient circuit can still be obtained, but with a “twisted” input-output relation (i.e., having x1 -> y3 and y1 -> x3 instead of x1 -> x3 and y1 -> y3. 

Fundamentally, there is no requirement for the input and the output to be aligned. However, considering the conditional nature of the computation (i.e., if the control qubit $q$ is in the state zero, the doubling does not occur and the value remains as $x_1$ instead of being transformed into $y_3$) and the circuit will be incorporated into a larger scheme of scalar multiplication, a direct alignment will be helpful for clarity of the operations, which can be done simply by appending (controlled) swap gates. 

Nevertheless, as previously described, the construction of point doubling may necessitate more space (i.e., ancilla registers) than that of point addition. While the latter, as proposed by Banegas et al. \cite{banegas2021concrete}, requires one ancilla register used as a placeholder for division operation (see Fig. \ref{fig:ecc_pa}), two ancilla registers will be required for performing point doubling. Below, we provide three schemes of point-doubling circuits to suit different implementation preferences.

% The proposed circuits for performing the point doubling are presented in Fig \ref{}. As shown, these three circuits consist of two input/output n-sized registers, one control qubit q, and two n-sized ancilla registers for containing intermediate results. Additionally, more multi-controlled gates are present throughout the circuit. This is due to the double scalar multiplication scenario, which necessitates the circuit to be conditional to the state of the control qubit (i.e., stay in its initial state (i.e., x1 and y1) when the control qubit is in the state 0). Additionally, note that our proposal lies in the high-level view of the circuit arrangement, whereas the underlying field operations and subroutines (e.g., multiplication, squaring, and so on) may use existing techniques, e.g., schoolbook, or Karatsuba multiplication proposed in [banegas, putranto] with relevant adjustments to the number of qubits used in the circuit. The state change corresponding to these circuits is presented in Table . In detail, the full steps (up to line 15) are for the third scenario, while it stops at line 10 and 12 for the second scenario (fig b), and the first scenario (fig a), respectively.

\begin{figure}[!htb] 
\begin{subfigure}{\textwidth}  
\centering
\resizebox{\linewidth}{!}{
\scalebox{1.5}{
\Qcircuit @C=1em @R=.7em {
&& \lstick{\ket{q}} & \qw & \qw & \ctrl{1} & \qw & \ctrl{2} & \ctrl{2} & \ctrl{2} & \ctrl{1} & \ctrl{1} &\push{\rule{0em}{1.2em}}\qw & \ctrl{1} & \qw & \qw & \qw &\rstick{\ket{q}}\qw & & & & & \\
&&\lstick{\ket{x_1}} & {{/}^{n}}\qw & \ctrl{1} & \ctrl{1} & \ctrl{2} & \qw & \qw & \qw  & \gate{S} & \targ & \qw & \qswap  & \qw & \qw& \push{\rule{0em}{1.2em}}\qw  &\rstick{\ket{x_3} or \ket{x_1}}\qw \\
&& \lstick{\ket{y_1}} & {{/}^{n}}\qw & \ctrl{1} & \gate{M} & \qw & \gate{S} & \targ & \gate{+a} & \ctrl{1} & \qw & \ctrl{1} & \qswap \link{-1}{0} & \push{\rule{0em}{1.2em}}\qw & \qw & \qw &\rstick{\ket{y_3} or \ket{y_1}}\qw \\
&& \lstick{\ket{0}} & {{/}^{n}}\qw & \gate{D} & \ctrl{-1} & \targ & \ctrl{-1} & \ctrl{-1} & \gate{+1} & \ctrl{1} & \qw & \ctrl{1} & \qw & \qw  & \qw & \push{\rule{0em}{1.2em}}\qw & \rstick{\ket{\lambda + 1}}\qw \\
&& \lstick{\ket{0}} & {{/}^{n}}\qw & \qw & \qw & \qw & \qw & \qw & \qw & \gate{M} & \ctrl{-3} & \gate{M} & \qw & \qw & \qw & \push{\rule{0em}{1.2em}}\qw &\rstick{\ket{0}}\qw}
}
}

\caption{Proposed point doubling circuit, with clearing one ancilla register}\label{fig:ecc_pd_balanced}
\vspace{1.2em}
\end{subfigure}
\begin{subfigure}{\textwidth}
\centering
\resizebox{0.9\linewidth}{!}{
\scalebox{1.5}{
\Qcircuit @C=1em @R=.7em {
&& \lstick{\ket{q}} & \qw & \qw & \ctrl{1} & \qw & \ctrl{2} & \ctrl{2} & \ctrl{2} & \ctrl{1} & \ctrl{1} & \ctrl{1} & \push{\rule{0em}{1.2em}}\qw &\rstick{\ket{q}}\qw & & & & & &  \\
&&\lstick{\ket{x_1}} & {{/}^{n}}\qw & \ctrl{1} & \ctrl{1} & \ctrl{2} & \qw & \qw & \push{\rule{0em}{1.2em}}\qw  & \gate{S} & \targ & \qswap & \qw  &\rstick{\ket{x_3} or \ket{x_1}}\qw \\
&& \lstick{\ket{y_1}} & {{/}^{n}}\qw & \ctrl{1} & \gate{M} & \qw & \gate{S} & \targ & \gate{+a} & \ctrl{1} & \qw  & \qswap \link{-1}{0}  &\push{\rule{0em}{1.2em}} \qw &\rstick{\ket{y_3} or \ket{y_1}}\qw \\
&& \lstick{\ket{0}} & {{/}^{n}}\qw & \gate{D} & \ctrl{-1} & \targ & \ctrl{-1} & \ctrl{-1} & \gate{+1} & \ctrl{1} & \qw & \qw  & \push{\rule{0em}{1.2em}}\qw & \rstick{\ket{\lambda+1}}\qw \\
&& \lstick{\ket{0}} & {{/}^{n}}\qw & \qw & \qw & \qw & \qw & \qw & \qw & \gate{M} & \ctrl{-3} & \qw & \push{\rule{0em}{1.2em}}\qw &\rstick{\ket{(\lambda +1)x_3} or}\qw \\
&&&&&&&&&&&&&&\rstick{\ket{(\lambda +1)y_1}}}
}
}
\caption{Alternative 1: Without uncomputation\label{fig:ecc_pd_nouncompute}}
\vspace{1.2em}
\end{subfigure}
\begin{subfigure}{\textwidth}  
\centering
\resizebox{\linewidth}{!}{
\scalebox{1.5}{
\Qcircuit @C=1em @R=.7em {
&& \lstick{\ket{q}} & \qw & \qw & \ctrl{1} & \qw & \ctrl{2} & \ctrl{2} & \ctrl{2} & \ctrl{1} & \ctrl{1} &\push{\rule{0em}{1em}}\qw & \ctrl{1} & \ctrlo{1} & \ctrlo{1} & \qw &\rstick{\ket{q}}\qw & & & & & \\
&&\lstick{\ket{x_1}} & {{/}^{n}}\qw & \ctrl{1} & \ctrl{1} & \ctrl{2} & \qw & \qw & \qw  & \gate{S} & \targ & \qw & \qswap  & \ctrl{2} & \ctrl{1} & \push{\rule{0em}{1em}}\qw  &\rstick{\ket{x_3} or \ket{x_1}}\qw \\
&& \lstick{\ket{y_1}} & {{/}^{n}}\qw & \ctrl{1} & \gate{M} & \qw & \gate{S} & \targ & \gate{+a} & \ctrl{1} & \qw & \ctrl{1} & \qswap \link{-1}{0} & \push{\rule{0em}{1em}}\qw & \ctrl{1} & \qw &\rstick{\ket{y_3} or \ket{y_1}}\qw \\
&& \lstick{\ket{0}} & {{/}^{n}}\qw & \gate{D} & \ctrl{-1} & \targ & \ctrl{-1} & \ctrl{-1} & \gate{+1} & \ctrl{1} & \qw & \ctrl{1} & \gate{+1} &\targ & \gate{D} & \push{\rule{0em}{1em}}\qw & \rstick{\ket{\lambda} or \ket{0}}\qw \\
&& \lstick{\ket{0}} & {{/}^{n}}\qw & \qw & \qw & \qw & \qw & \qw & \qw & \gate{M} & \ctrl{-3} & \gate{M} & \qw & \qw & \qw & \push{\rule{0em}{1em}}\qw &\rstick{\ket{0}}\qw}
}
}
\caption{Alternative 2: Fully uncompute when $q = 0$, otherwise leaving one ancilla as $\lambda$}\label{fig:ecc_pd_full}
\vspace{1.2em}
\end{subfigure}
\caption{Our proposed point doubling circuits for binary elliptic curves: (a) balanced version that clears one ancilla registers, and two alternatives of (b) without uncomputation for lower depth and lower gate count, and (c) full uncomputation when control qubit $q = 0$, and with a garbage ancilla in state $\lambda$ when $q = 1$.} \label{fig:ecc_all_pd}
\end{figure}
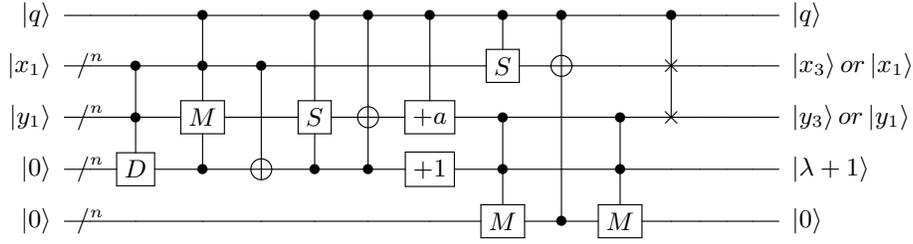
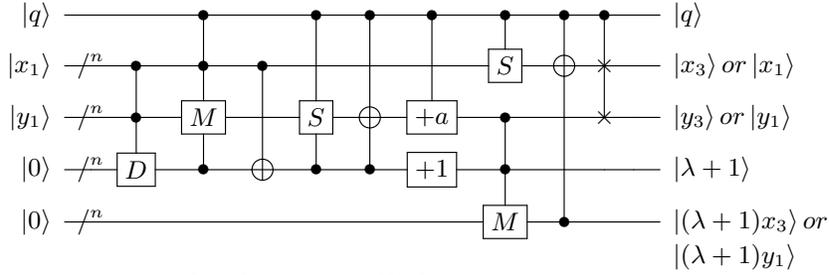
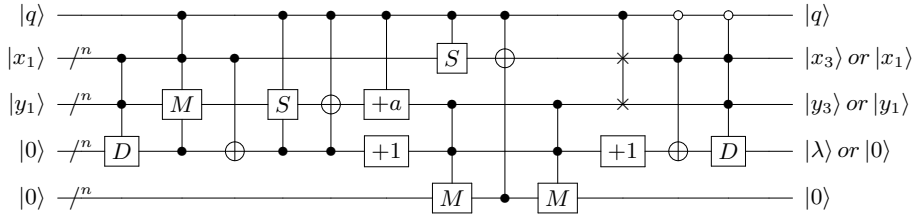

The proposed circuits for performing point doubling are illustrated in Fig. \ref{fig:ecc_all_pd}. These circuits consist of two $n$-sized input/output registers, a control qubit $q$, and two $n$-sized ancilla registers to store intermediate results. Additionally, the presence of multiple multi-controlled gates throughout the circuit results from the circuit's conditional nature, wherein it remains in the initial state (i.e., $x_1$ and $y_1$) when the control qubit is in the state $\ket{0}$. It is important to highlight that our proposal focuses on the high-level structure of the circuit arrangement, whereas the underlying field operations and subroutines (e.g., multiplication, squaring) may employ existing techniques such as Schoolbook or Karatsuba multiplication as proposed in \cite{banegas2021concrete,putranto2023depth,jang2022optimized}, with necessary adjustments made to accommodate the number of qubits required on each construction. The state change corresponding to these circuits is presented in Table \ref{tab:pd_statechange}. In detail, the complete steps (up to line 15) are for the third scenario (Fig. \ref{fig:ecc_pd_full}), while the second scenario (Fig. \ref{fig:ecc_pd_nouncompute}) and the first scenario (Fig. \ref{fig:ecc_pd_balanced}) terminate at lines 10 and 12, respectively.

\begin{table}[]
% \large
% \normalsize
\centering
\caption{Point Doubling State Change}
\label{tab:pd_statechange}
\begin{tabular}{|c|c|c|}
\hline
% \rowcolor{Gray}
\textbf{Step} & \textbf{$q = 1$}                          & \textbf{$q = 0$}              \\ \hline
1             & $anc_1 = \frac{y_1}{x_1}$               & $anc_1 = \frac{y_1}{x_1}$   \\
2             & $y = 0$                                 & $y = y_1$                   \\
3  & $anc_1 = \frac{y_1}{x_1} + x_1 = \lambda$ & $anc_1 = \frac{y_1}{x_1} + x_1 = \lambda$    \\
4             & $y = \lambda^2$                         & $y = y_1$                   \\
5             & $y = \lambda^2 + \lambda$               & $y = y_1$                   \\
6             & \cellcolor{LightCyan} $y = \lambda^2 + \lambda + a = x_3$     & $y = y_1$                   \\
7             & $anc_1 = \lambda + 1$                   & $anc_1 = \lambda + 1$       \\
8             & $anc_2 = (\lambda + 1) x_3$             & $anc_2 = (\lambda + 1) y_1$ \\
9             & $x = {x_1}^2$                           & $x = x_1$                   \\
10            & \cellcolor{LightCyan} $x = {x_1}^2 + (\lambda + 1) x_3 = y_3$ & $x = x_1$                   \\
11            & $anc_2 = 0$                             & $anc_2 = 0$                 \\

12            & \cellcolor{LightCyan} $swap: x = x_3, y=y_3$                  & \cellcolor{LightCyan} $none: x = x_1, y=y_1$      \\
13            & $anc_1 = (\lambda+1) -1 = \lambda$      & $anc_1 = \lambda$           \\
\rowcolor{Gray}
14 & $anc_1 = \lambda$                         & $anc_1 = \lambda - x_1 =    \frac{y_1}{x_1}$ \\
\rowcolor{Gray}
15            & $anc_1 = \lambda$                       & $anc_1 = 0$  \\ \hline              
\end{tabular}
\end{table}

In the first circuit (Fig. \ref{fig:ecc_pd_balanced}), we propose a balanced approach that strikes a tradeoff between the number of operations and the need to clear the ancilla qubit. While this circuit involves a relatively smaller number of operations, it requires an additional multiplication circuit for performing uncomputation upon one of the ancilla registers. As a result, we obtain one cleared ancilla register that can be used for subsequent computations, and one dirty (i.e., does not revert to its initial state after use) ancilla register in the state ($\lambda+1$). This is our favored version because we can secure one clean register with relatively minimal effort.

Alternatively, if circuit depth and gate count take precedence over qubit count, the more suitable circuit would be as illustrated in Fig. \ref{fig:ecc_pd_nouncompute}. Here, the circuit only performs the expected point doubling operation without considering any uncomputation for ancilla registers. This minimizes depth and the number of subroutines, but we are left with two dirty ancilla registers. 

Regarding the third case, our initial goal was to clear all ancilla registers. However, we have not found a more efficient method to fully uncompute them for all possible states of the control qubit ($\ket{0}$ or $\ket{1}$). A complete uncomputation can be achieved when $q=0$, but the state of $\lambda$ remains a dirty ancilla when $q=1$. Note that $\lambda$ from the previous state (i.e., $x_1 + \frac{y_1}{x_1}$) may not be in the same value as $\lambda$ in the subsequent operation (i.e., $x_3 + \frac{y_3}{x_3}$); due to this potential differences in value, we should not uncompute it by utilizing $x_3$ and $y_3$ when $q=1$. Had it been the same, it would allow us to obtain two perfectly-uncomputed, clean ancilla registers. This can be done by appending another controlled multiplication circuit and Toffoli gate targeting that ancilla register. 

Nevertheless, this third construction is still useful; Evidently, at the time when $q=0$ indicates that the point doubling does not occur \textemdash meaning that most likely a point addition is taking place. This register can be repurposed using a clever arrangement to substitute the ancilla register in the point addition circuit (i.e., the one for performing division operation in Fig. \ref{fig:ecc_pa}). The uncomputation itself can be performed by appending a series of Toffoli gates, a multi-controlled and negative-controlled division operation, and an addition by a constant. This circuit serves as a beneficial alternative construction when circuit width or qubit utilization is the most prioritized quantum resource.

The high-level resource cost of the proposed circuits can be summarized as follows. Compared to point addition, point doubling construction employs one more ancilla register and significantly more controlled and multi-controlled operations due to its non-fixed second point. In detail, for the first scenario (balanced), a total of one division, three multiplications (including multi-controlled version), and two squarings (one multi-controlled) are employed, one controlled swap (i.e., Fredkin gate), as well as several (controlled and multi-controlled) additions, with one clean ancilla registers and one dirty ancilla register. For the second scenario, one less multiplication is employed, with the tradeoff of having both ancilla registers dirty. Additionally, regarding the third alternative of having two clean ancilla registers when the control qubit state is zero, it requires additional subroutines of one negative-control Toffoli series (elevated control from addition via CNOT gates), one negative multi-controlled division, and one addition by a constant. Note that we do not elaborate further on the exact resource since the presence of multi-controlled operations requires a more complex circuit decomposition. Nevertheless, we plan to investigate it further on a quantum simulator in our future work to obtain a more concrete resource estimation.

\section{Discussions and Limitations}
In this study, we delve into the topic of elliptic curve point doubling circuits, which has yet to be further examined in the literature. After presenting the design and description of our approach in the previous section, we now provide discussions related to the broader implications of our proposal.

\textbf{Transforming to prime curves.} We begin our study from the binary elliptic curves, which are relatively simpler than prime curves. In the case of prime curves, the quantum circuit will be more complicated because it cannot make use of the simplicity of field operation in binary curves. For instance, an addition in the prime fields requires a full adder, whereas binary fields only necessitate one Toffoli gate for each bit. Additionally, FLT-based inversion, which is comparable in performance to GCD-based inversion in quantum binary elliptic curves, has also not been considered to date for its use on prime curves due to the high resource requirements. Similarly, squaring operations are favorable in the binary case due to their relatively efficient construction (i.e., by leveraging a simple LUP decomposition), which are not applicable to prime curves. Even though there is an advantage in prime curves in terms of intuitive verification due to their nature of resembling decimal calculation, it requires more space and operations that are arguably more complicated and resource-intensive.

\textbf{Relation to scalar multiplication.} To realize a quantum elliptic curve scalar multiplication, existing methods rely upon a series of point addition circuits as the sole components. Therefore, the computation is in the form of $Q = kP = P+P+…+P$ for $k$ times. Considering a more general implementation without limiting its use to Shor’s algorithm, performing scalar multiplication by incorporating point doubling alongside point addition can potentially reduce the depth of the circuit and the number of operations. Moreover, the availability of designs for both point addition and point doubling opens up the opportunity to explore various classical elliptic curve point/scalar multiplication (ECPM) techniques (e.g., signed digit method, M-ary method) to be explored in the quantum realm in search of more efficient circuits.  

\textbf{For use in Shor’s ECDLP.} In order to create a more theoretically accurate and complete Shor’s ECDLP circuit, point doubling will need to be integrated into the existing double scalar multiplication (that currently consists entirely of point addition subroutines) to cover the cases when the doubling of points occur. Note that for this algorithm, the input comes from the Walsh-Hadamard, so that the circuit is expected to be able to compute all possible cases (i.e., any combination of zero and one within the circuit). More importantly, the constant value $k$ is unknown. For this reason, a more thorough conditional mechanism is required to control whether point doubling or point addition is in effect during the certain computation phase, resulting in more complex multi-control operations on the circuit. Additionally, in scenarios where both points of interest are quantum values, a comparator circuit may need to be employed to determine whether both values are identical. Note that for the specific use in Shor’s ECDLP, both point addition and point doubling may need to be employed altogether to cover all cases, and other implementation requirements (e.g., unique representation for history independence \cite{haner2020improved}, uncomputing garbage outputs to prevent unwanted interference \cite{orts2020review}) will need to be taken into account, which will be explored in our future work.

\textbf{Limitations.} Even though our work has provided an initial step to explore further into point doubling, there are still various aspects that require further investigation. This includes how to correctly integrate it with the point addition circuit and whether the previous assumptions taken for Shor’s ECDLP regarding the double-scalar multiplication still stand in this case, which is an interesting research problem. %, especially for the specific use in Shor’s algorithm, 
We leave these topics for our future work.

%%%%%%%%%%%%%%%%%%%%%%%%%%%%%%%%%%%%%%%%%%%%%%%%%%
\section{Conclusions and Future Work}
In this study, we have examined the point-doubling operation for binary elliptic curves, which are required in the stricter case of Shor’s algorithm. We began by analyzing the point doubling formula, identifying the inherent challenges in its construction, and presenting a possible solution. Subsequently, we designed quantum circuits for elliptic curve point doubling to cater to different scenarios, which shows the need for one more ancilla register compared to point addition, and while they may be comparable in terms of the number of subroutines, more complex multi-controlled operations are required than that of point addition. In addition, we provide a more in-depth discussion of the implications and concerns in incorporating the circuit into Shor’s algorithm. To obtain a more detailed resource estimation for point doubling and complete double scalar multiplication for Shor’s algorithm, we plan to construct the circuit in the existing quantum computing simulators and run the resource analysis as our future work. 

%%%%%%%%%%%%%%%%%%%%%%%%%%%%%%%%%%%%%%%%%%%%%%%%%%

% \nocite{*}
% ================
\bibliographystyle{splncs04}
% \printbibliography
\typeout{}
\bibliography{bib}
% ===============

%

% \begingroup
% % \setstretch{0.8}
% \setlength\bibitemsep{0pt}
% \printbibliography
% \endgroup

\end{document}